\documentclass[a4paper,fleqn]{cas-sc}
\usepackage[numbers]{natbib}
\usepackage{placeins,caption} 
\usepackage{xcolor}
\begin{document}
\let\WriteBookmarks\relax
\def\floatpagepagefraction{1}
\def\textpagefraction{.001}
\shorttitle{Split-Stagger Starts} 
\shortauthors{Souaiaia}
%\begin{frontmatter} ORCID

\title [mode = title]{Estimating the Time Advantage of Split-Staggered\\Starts in Record-Breaking Middle-Distance Races} 
\date{September 8, 2026}

%  we 

%\author[1]{Tade Souaiaia}[type=editor,auid=000,bioid=1,orcid=33]
\author[1]{Tade Souaiaia}[orcid=0000-0003-3922-1372] 
\cormark[1]
%\fnmark[1]  % Highlights
\ead{tade.souaiaia@downstate.edu} 
\credit{conceptualisation, investigation, formal analysis, methodology, writing -original draft}
\address[1]{Department of Cell Biology, SUNY Downstate Health Sciences, Brooklyn, NY, USA} 
\cortext[cor2]{Corresponding author}

\begin{abstract}
In July 2026, Josh Kerr broke the 26-year-old mile world record, the latest in an unusual concentration of 
recent middle-distance records that has led to widespread speculation about the role of advancing shoe and other technologies.
However, like almost every other recent record-breaking performance, 
Kerr's race featured a split-staggered start---an often overlooked, relatively novel start configuration 
that places a subset of athletes in a front group in the outside lanes, reducing the geometric burden 
associated with accelerating on tight turns while also avoiding the bulk of the extra distance 
associated with outside positioning in a traditional distance race.

Here, these advantages are quantified using empirical lane-performance data, OMEGA
measurements of realized race distance, video reconstruction, and a force-based
model of curved running. Applying the resulting corrections to record breaking performances
from 2026, estimated advantages range from approximately 0.1~s to more than
1~s, depending on the event, race configuration, and model assumptions.
Although small in absolute terms, these advantages demonstrate that start 
configuration represents a meaningful and previously overlooked source 
of performance advantage in modern middle-distance racing---large enough to 
make the difference between a world record and an otherwise great performance.
\end{abstract}
\begin{keywords}
Athletics \sep Physics \sep Modeling \sep Running \sep Geometry \sep Performance
\end{keywords}
\maketitle
\section*{Introduction}
\label{sec:introduction}
Middle-distance records are falling at a remarkable rate.
Of the 24 men's middle-distance world-records broken since 2000,
14 occurred between 2022 and 2026 ($p \approx 1.7\times10^{-5}$; Appendix: ~\nameref{asec:records}), fueling
widespread speculation that advancing shoe technology or other equipment changes are responsible \cite{kerr_shoes, kerr_suit, kerr_chamber}.
Despite the popularity of these explanations \cite{mar_shoes, bicarb, michael_johnson},
direct evidence that shoe technology provides a measurable advantage in middle-distance track racing remains limited
\cite{no_shoes}. Here the focus is on an under-examined feature especially associated with
the recent run of record-breaking performances: a lane draw placing the winning
athlete in the front group of a split-staggered start.

Unlike the traditional "waterfall" start, in which all athletes start behind
a single curved starting line and are immediately free to move toward the inside rail,
a split-staggered start divides the field into two groups, with a subset (approximately one-third)
starting ahead and running a portion of the race at a wider radius before merging 
with the other athletes along the rail.  The forward stagger is calculated to compensate
for this wider path and produce a fair race, such that athletes in both groups following 
their inside line will have covered an identical distance when they merge.

Despite this, all 14 world records set under this format since 2016 have been
won from the forward group (Appendix: ~\nameref{asec:records}).
Under the deliberately naive null that potential record setters are randomly distributed
between groups, the probability of this is roughly one in five million.
One of the more notable recent indoor waterfall performances came when Cole 
Hocker set an American record and came within three seconds of
the world record in the indoor 2000\,m \cite{FlashResults2026Hokie2000m}.
This record was short-lived, however: the next day Hobbs Kessler (an athlete with slower 1500\,m and mile
personal bests than Hocker) broke both the American and world records in the 2000m, running more than 4 seconds faster than
Hocker from the front group in a two-turn split-staggered start \cite{Monti2026HoeyKessler}.

The pattern extends beyond which group wins split-staggered races.
Excluding the 800m, which is started in lanes, only 2 of the 24 
men's middle-distance world-records since 2000 were set from a traditional waterfall
start in the last ten years: Jakob Ingebrigtsen's outdoor 2000m and 2-mile records, both in 2023.
The timing of this pattern coincides with a rapid evolution in start configurations.
Before 2010, the "waterfall" start was the norm for both indoor and outdoor
middle-distance races. From approximately 2015 through 2025, one-turn split-staggered starts became
the dominant indoor format, while outdoor races largely remained traditional.
In 2026, the format changed again, with two-turn split staggers appearing indoors
and one-turn split staggers outdoors; these new configurations were used in each 
of the six world or area record performances seen here (Table~\ref{tab:2026_intro}).
\begin{table}[h]
\centering
\small
\caption{Record Breaking Middle-Distance Performances (2026).} 
\label{tab:2026_intro}
\begin{tabular}{llllll}
\toprule Event & Result & Time & Athlete & Month & Configuration \\ \midrule
800\,m Indoor  & World Record & 1:42.50 & Josh Hoey & Jan 26 & Two-Turn Lane Stagger \\
2000\,m Indoor  & World Record & 4:48.79 & Hobbs Kessler & Jan 26 & Two-Turn Split-Staggered \\
1000\,m Indoor  & European Record & 2:14.52 & Mohamed Attaoui & Feb 26 & Two-Turn Split-Staggered \\
Mile\, Indoor   & American Record & 3:45.94 & Cole Hocker & Feb 26 & Two-Turn Split-Staggered\\
1000\,m Outdoor & World Record & 2:11.83 & Emmanuel Wanyonyi & July 26 & One-Turn Split-Staggered Start\\
Mile Outdoor & World Record & 3:42.66 & Josh Kerr & July 26 & One-Turn Split-Staggered Start \\ \bottomrule
    \vspace{-0.5cm} 
\end{tabular} \end{table}
\FloatBarrier
At first glance, it may appear puzzling why there is such a clear enrichment
of world records for a start configuration that does not change the race distance.
The source of this advantage becomes clearer, however, when one considers the tactical 
decisions faced by middle-distance athletes under a traditional start in events that begin immediately before a turn (e.g., 1000m, mile, or 2000m).
In isolation, an inside start position is preferable because it allows one to run
the shortest possible path along the rail. In a real race, however, this comes with a substantial cost.
The athlete must either \textbf{(i)} fight to maintain the rail against multiple runners accelerating
around the outside, intermittently accelerating along a tighter turning radius to prevent them from reaching
the rail, or \textbf{(ii)} accelerate conservatively while accepting a position behind the athletes
most determined to reach the front. Both strategies have serious drawbacks.
Strategy \textbf{(i)} requires greater energy expenditure and exposes the athlete to contact, collisions, and falls.
Strategy \textbf{(ii)} minimizes both distance and early energy expenditure, but leaves the athlete likely to be
"boxed in" behind competitors who must eventually be overtaken, requiring additional energy and distance
later in the race when fatigue is greater.

\FloatBarrier \begin{figure}[!h] \centering
\caption{\scriptsize{
        Optimal paths under traditional waterfall and split-stagger  starts.
        The optimal waterfall path exposes the athlete to the tightest turning
        radius, while the one-turn split stagger omits a portion
        of the first turn and allows the remainder to be run at a wider radius.
        Two-turn split staggers extend this advantage. 
        Indoors, different start formats also allow athletes to avoid
        running up one of the banked turns, providing further mechanical advantage.
}}
\includegraphics[trim=1.5cm 0.85cm 1.5cm 0.4cm, clip, height=6.4cm]{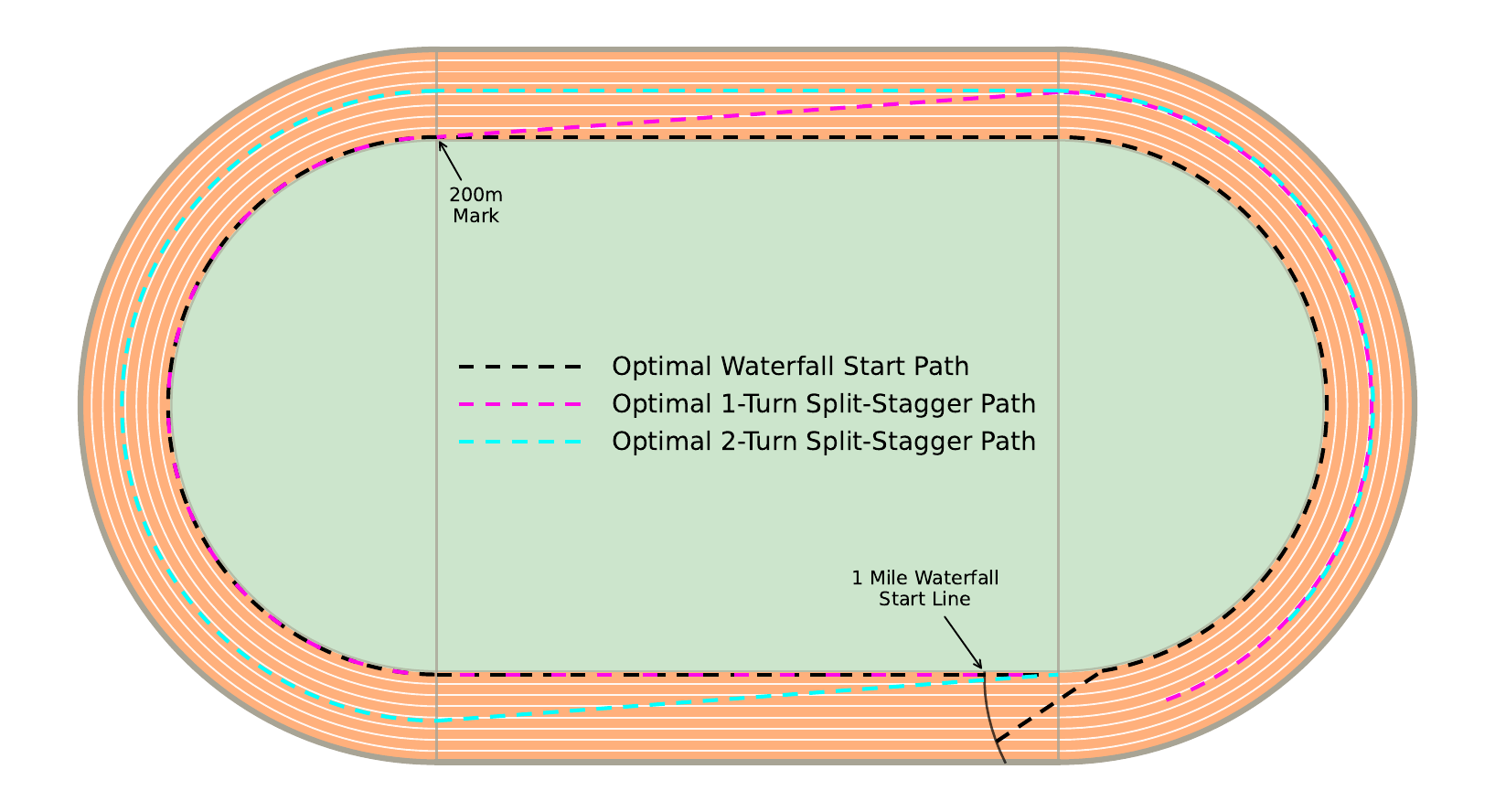}
\label{fig:optimal_paths} \end{figure} \FloatBarrier

It is for these reasons that athletes attempting world records frequently prefer an outside starting position,
where they can accelerate smoothly, avoid the early fight for the rail, run around competitors when necessary,
and choose when to move in. That athletes accept the obvious cost of running several extra meters to obtain these advantages can be seen
in broadcast video of the most recent prominent attempts on the mile world record before Kerr. In these races, athletes started in Lane 8
(Ingebrigtsen, 2023 \cite{ingebrigtsen2023mileattempt}), Lane 8 (El Guerrouj, 2000 \cite{crystalpalace}),
Lane 7 (El Guerrouj, 1999 \cite{rome}), Lane 4 (Morceli, 1993 \cite{morceli1993mile}), and Lane 7 (Cram, 1985 \cite{cram1985mile}).

The split-staggered start changes this tradeoff.
The forward displacement of the outer starting positions compensates for the longer outside path,
allowing athletes to exploit the wider radius and greater freedom of movement without paying for it
in extra distance. This advantage arises through two physically distinct factors.

The first factor is geometric: running a wider curve at a fixed speed requires less centripetal
force and a smaller directional change per meter than running a tight one,
with the split-stagger formats progressively reducing exposure to the tightest turns
(Figure~\ref{fig:optimal_paths}).
Indoors, banking adds another component to this advantage because lane position changes both
turning radius and the banking experienced by the athlete.
Because the official race distance is identical under both starting configurations, this component
reflects how the same distance is negotiated rather than how far the athlete runs.

\begin{figure}[!h] \centering 
\caption{\footnotesize{
Reconstructed one-lap paths for three elite mile performances under
traditional waterfall starts, compared with the optimal waterfall path.
An unobstructed athlete can reach the rail approximately 15--30~m into
the opening bend, whereas all three athletes remained outside the optimal
path substantially longer.  The resulting excess distances were approximately 5.5~m for Myers
(OMEGA), 4.6~m for Ingebrigtsen (OMEGA), and 6~m for El Guerrouj (video estimate).}}
\includegraphics[trim=1.5cm 0.85cm 1.5cm 0.4cm, clip, height=6.4cm]{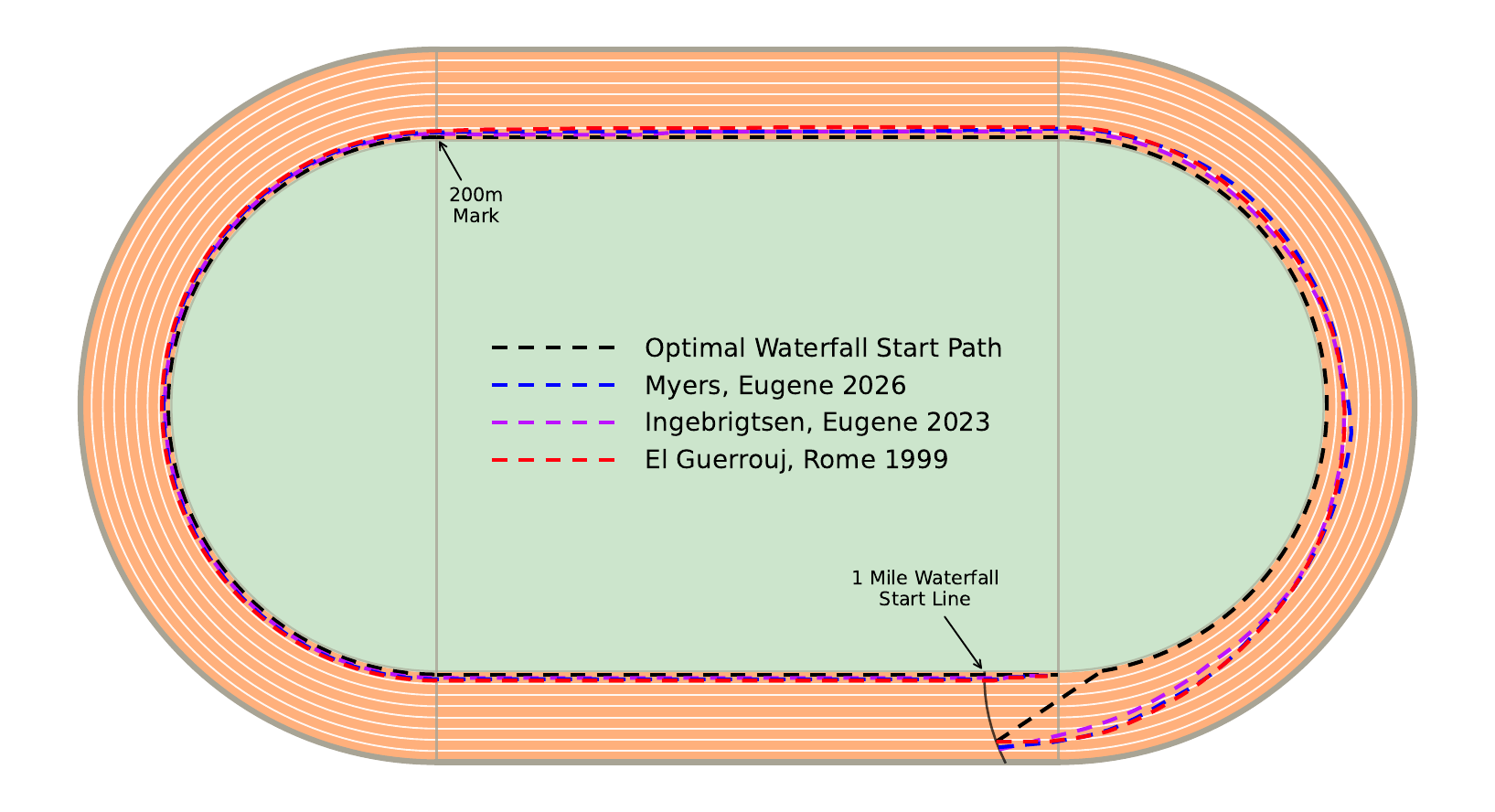}
\label{fig:real_paths} \end{figure}

The second factor is the actual distance traveled.
Although the split stagger is designed so that the optimal paths cover the same distance,
athletes under a traditional waterfall start rarely follow the shortest possible path.
For an athlete starting outside, doing so would require an almost immediate move to the rail,
which is generally impossible in a crowded field without colliding with other athletes
(Figure~\ref{fig:real_paths}).
The athlete therefore runs additional distance before reaching the rail.
We quantify this distance directly from OMEGA athlete-tracking data where available and reconstruct
it from race video for older performances.

Combining these two factors allows us to estimate the advantage provided by the split-staggered
start and ask the central counterfactual question: would these records still have been broken
under traditional starting conditions?

\FloatBarrier  \FloatBarrier

%%%%%%%%%%%%%%%%%%%%%%%%%%%%%%%%%%%%%%%%%%%%%%%%%%%%%%%%%%%%%%%%%%%%%%%%%%%%%%%%%%%%%%%%%%%%%%%  
%%%%%%%%%%%%%%%%%%%%%%%%%%%    METHODS   %%%%%%%%%%%%%%%%%%%%%%%%%%%%%%%%%%%%%%%%%%%%%%%%%%%%%%  
%%%%%%%%%%%%%%%%%%%%%%%%%%%%%%%%%%%%%%%%%%%%%%%%%%%%%%%%%%%%%%%%%%%%%%%%%%%%%%%%%%%%%%%%%%%%%%%  

\clearpage  

% software

\section*{Methods} 
\subsection*{Factor 1: Wider Turn Radius}
\label{sec:factor1}
\subsubsection*{Speed-Independent Empirical Calibration}
No published lane-time data are available for middle-distance races, which are typically
run in lane 1, so lane effects are calibrated using published outdoor and indoor 200m sprint data.
For outdoor races, the pooled linear lane effect reported by \citet{munro2022} are used as the primary empirical calibration.
Munro estimated a 0.0189s improvement in 200m time for each lane outward, an estimate that is further supported
by the independent analysis of \citet{chen2022blog} (0.018s per lane).
For a target athlete with effective lane position $\ell_T$ and comparison
athlete with effective lane position $\ell_C$, we can therefore estimate the
outdoor geometric advantage as
\[ \Delta t_{\mathrm{geom,out}} = 0.0189(\ell_T-\ell_C). \]
\citet{munro2022} found no significant difference in the lane effect between male and female
sprinters despite their different race speeds, providing some empirical support
for a speed-independent calibration. For indoor races, the 200m lane comparisons reported by
\citet{flanders_isbs} are used: lane 6 was 0.23s faster than lane 2 and
0.10s faster than lane 4. For a semicircular turn,
\begin{equation} \int \kappa^2\,ds = \int \frac{1}{r^2}\,ds = \frac{\pi}{r}. \end{equation}
Because the common factor $\pi$ is absorbed into the empirical calibration,
we defined the indoor radius burden for a set of turns as
\begin{equation} B_1 = \sum_q \frac{1}{r_q}.\label{eq:B1} \end{equation}
For the two-turn 200m calibration races, the corresponding lane contrasts are
\begin{equation} \Delta B_1(a,b) = 2\left(\frac{1}{r_a}-\frac{1}{r_b}\right). \end{equation}
A single proportionality constant was estimated by regression through the
origin:
\begin{equation} c_{\mathrm{in}} = \frac{0.23\,\Delta B_1(2,6) + 0.10\,\Delta B_1(4,6)}{\Delta B_1(2,6)^2+\Delta B_1(4,6)^2}.\end{equation}

Using the 1.00m lane width of the Flanders Expo calibration track gives
$c_{\mathrm{in}}=11.89\,\mathrm{s\,m}$. For the 2026 indoor race venues,
lane radii were calculated using the measured 1.07m lane width (see Table \ref{atab:constants}).
For a target and comparison athlete, the corresponding indoor geometric
advantage was therefore
\begin{equation} \Delta t_{\mathrm{geom,in}} = c_{\mathrm{in}}\sum_q\left(\frac{1}{r_{C,q}}-\frac{1}{r_{T,q}}\right). \end{equation}
For 1.07m race-lane width, this corresponds to an estimated lane 1--lane 4
difference of 0.211s over two turns. For the indoor 800m comparison, only one
differential turn is credited; the corresponding lane 1--lane 5 estimate is
0.134s.

\subsubsection*{Speed-Dependent Force-Demand Model}
\label{sec:model}
For a curved segment, the corresponding curvature exposure is
\begin{equation} \int \kappa(s)^2\,ds. \end{equation}
For a semicircular turn of radius $r$, this equals $\pi/r$ and is therefore
proportional to the radius burden $B_1$ defined above.
Following the centripetal-force framework used in bend-running biomechanics
\citep{chang1999,chang2007,churchill2016}, resultant support/propulsive demand
per unit mass was modeled as
\begin{equation} A_{\text{straight}} = \sqrt{g^2+a_t^2}, \qquad A_{\text{curve}} = \sqrt{g^2+a_t^2+v^4\kappa^2}. \end{equation}
Tangential acceleration $a_t$ was set to $0$, corresponding to constant speed
through the turn. The resulting increase in force demand is therefore
\begin{equation} A_{\text{curve}}-A_{\text{straight}} = g\left[\sqrt{1+\left(\frac{v^2\kappa}{g}\right)^2}-1\right]. \end{equation}
Expanding in $y=(v^2\kappa/g)^2$ gives
$\sqrt{1+y}-1 \approx y/2$, so integrating the force-demand increment over
time gives a curvature-related demand proportional to
\begin{equation} B_2 \propto B_1\frac{v^3}{2g}. \label{eq:B2} \end{equation}
Thus, for a fixed lane contrast, the model predicts that the time effect scales
approximately with the cube of running velocity. At the most extreme indoor
calibration condition ($v=9.09$m/s, lane 1), $v^2/(gr)\approx0.48$, corresponding
to $y\approx0.23$; at this value the first-order expansion overestimates the exact
force-demand increment by approximately 6\%. Because the geometric component
is generally below 0.15s in the races considered here, the resulting effect on
the total race corrections is small.

\subsubsection*{Calibration}
The speed-dependent estimate for each race was obtained by rescaling the corresponding speed-independent
empirical lane-time estimate by
\begin{equation} \left(\frac{v_{\mathrm{athlete}}}{v_{\mathrm{calib}}}\right)^3. \end{equation}
We used $v_{\mathrm{calib}}=9.73$ m/s outdoors, corresponding to the mean
200m time of 20.55s across 425 Diamond League performances
\citep{chen2022blog}, and $v_{\mathrm{calib}}=9.09$ m/s indoors,
corresponding to the mean recorded 200m times in lanes 2, 4, and 6 at
Flanders Expo (22.13s, 22.00s, and 21.90s)
\citep{flanders_isbs}. Athlete velocity was calculated from race distance and
official finishing time.

\subsection*{Factor 2: Realized Race Distance}
\label{sec:factor2}

Realized race distance was obtained for all six 2026 record breaking runs and for relevant 
"comparator" performances - athletes making credible record attempts under traditional 
waterfall starts or back-group configuration. 
Where available, actual distance traveled and excess distance were taken from OMEGA graphical race-analysis
reports in the official Wanda Diamond League race-analysis reports available through \texttt{omegatiming.com}
\citep{OmegaBowerman2023,OmegaBowerman2026,omega_london2026,OmegaMonaco1000m2026}.
For races without OMEGA distance data, path length was estimated from broadcast video.
Athlete lane position was recorded at 0.25s intervals over each curve until the athlete
reached lane 1. These observations were used to calculate the mean lane position for each
curve and the corresponding path length from the track geometry described above.
Effective lane was defined as the athlete's mean lane position over the portion of the race
impacted by starting configuration: the opening turn for one-turn configurations and
the opening two turns for two-turn configurations. For waterfall and back-group comparison
performances, effective lane was measured over the corresponding opening portion of the race.
Effective lane was used for the geometric correction in Factor~1, whereas excess distance
was calculated over the full race and used for the path-length correction described here.

The video method was checked against races with OMEGA Real Distance measurements and agreement was close:
across the four races with both measurements, the mean absolute difference was 0.15m (Appendix Table~\ref{atab:path_inputs}).
Therefore, it was used as an approximate reconstruction method for races without direct tracking data.
Hoey's 800m comparison was treated as geometry-only because the athletes were lane-locked through the relevant bend.
The path-length advantage for each comparison was calculated as
\begin{equation} \Delta D_{\mathrm{path}} = D_{\mathrm{extra,comparison}}-D_{\mathrm{extra,target}} \end{equation}
and converted to time using
\begin{equation} \Delta t_{\mathrm{path}} = \frac{\Delta D_{\mathrm{path}}}{v}  \end{equation}
where $v$ is the athlete's average race velocity.

\subsection*{Combining Components}
\label{sec:combining}

Factor 1 and Factor 2 estimate two distinct costs: the geometric cost of running
a tighter turn and the additional time required to cover a longer path.
For each comparison, the geometric term was calculated using the target and
comparison athletes' effective lane positions, while the path-length term was
calculated from the difference in their excess race distances.
The total estimated advantage was therefore
\begin{equation} \Delta t_{\mathrm{total}} = \underbrace{\Delta t_{\mathrm{geom}}(\ell_C,\ell_T)}_{\text{Factor 1}} + \underbrace{\frac{D_{\mathrm{extra,comparison}}-D_{\mathrm{extra,target}}}{v}}_{\text{Factor 2}}. \label{eq:combined} \end{equation}
where $\ell_C$ and $\ell_T$ are the effective lane positions of the comparison
and target athletes, respectively, and $v$ is the target athlete's average
race velocity. The geometric term was evaluated using both the speed-independent
empirical estimate and the speed-dependent model, producing two corresponding
estimates of the total advantage.

%%%%%%%%%%%%%%%%%%%%%%%%%%%%%%%%%%%%%%%%%%%%%%%%%%%%%%%%%%%%%%%%%%%%%%%%%%%%%%%%%%%%%%%%%%%%%%%  
%%%%%%%%%%%%%%%%%%%%%%%%%%%%%%%%%%%%%%%%%%%%%%%%%%%%%%%%%%%%%%%%%%%%%%%%%%%%%%%%%%%%%%%%%%%%%%%  
%%%%%%%%%%%%%%%%%%%%%%%%%%%%%%%%%%%%%%%%%%%%%%%%%%%%%%%%%%%%%%%%%%%%%%%%%%%%%%%%%%%%%%%%%%%%%%%  
%%%%%%%%%%%%%%%%%%%%%%%%%%%    RESULTS   %%%%%%%%%%%%%%%%%%%%%%%%%%%%%%%%%%%%%%%%%%%%%%%%%%%%%%  
%%%%%%%%%%%%%%%%%%%%%%%%%%%    RESULTS   %%%%%%%%%%%%%%%%%%%%%%%%%%%%%%%%%%%%%%%%%%%%%%%%%%%%%%  
%%%%%%%%%%%%%%%%%%%%%%%%%%%    RESULTS   %%%%%%%%%%%%%%%%%%%%%%%%%%%%%%%%%%%%%%%%%%%%%%%%%%%%%%  
%%%%%%%%%%%%%%%%%%%%%%%%%%%    RESULTS   %%%%%%%%%%%%%%%%%%%%%%%%%%%%%%%%%%%%%%%%%%%%%%%%%%%%%%  
%%%%%%%%%%%%%%%%%%%%%%%%%%%    RESULTS   %%%%%%%%%%%%%%%%%%%%%%%%%%%%%%%%%%%%%%%%%%%%%%%%%%%%%%  
%%%%%%%%%%%%%%%%%%%%%%%%%%%    RESULTS   %%%%%%%%%%%%%%%%%%%%%%%%%%%%%%%%%%%%%%%%%%%%%%%%%%%%%%  
%%%%%%%%%%%%%%%%%%%%%%%%%%%    RESULTS   %%%%%%%%%%%%%%%%%%%%%%%%%%%%%%%%%%%%%%%%%%%%%%%%%%%%%%  
%%%%%%%%%%%%%%%%%%%%%%%%%%%%%%%%%%%%%%%%%%%%%%%%%%%%%%%%%%%%%%%%%%%%%%%%%%%%%%%%%%%%%%%%%%%%%%%  
%%%%%%%%%%%%%%%%%%%%%%%%%%%%%%%%%%%%%%%%%%%%%%%%%%%%%%%%%%%%%%%%%%%%%%%%%%%%%%%%%%%%%%%%%%%%%%%  
%%%%%%%%%%%%%%%%%%%%%%%%%%%%%%%%%%%%%%%%%%%%%%%%%%%%%%%%%%%%%%%%%%%%%%%%%%%%%%%%%%%%%%%%%%%%%%%  
%%%%%%%%%%%%%%%%%%%%%%%%%%%%%%%%%%%%%%%%%%%%%%%%%%%%%%%%%%%%%%%%%%%%%%%%%%%%%%%%%%%%%%%%%%%%%%%  

\vspace{0.5cm} 
\section*{Results}
\label{sec:results}
\begin{center}
\footnotesize
\setlength{\tabcolsep}{3.5pt}
\renewcommand{\arraystretch}{1.15}

\captionof{table}{Estimated effect of the starting configuration on the 2026 record
performances. Advantage ranges include the speed-independent and
speed-dependent estimates and, where applicable, the range across the
historical comparisons considered.}
\label{tab:combined_results}

\begin{tabular}{
>{\raggedright\arraybackslash}p{3.0cm}
c
>{\raggedright\arraybackslash}p{3.4cm}
c
c
>{\raggedright\arraybackslash}p{2.0cm}
}
\toprule
2026 Performance & Time & Comparisons considered & Advantage (s) & Counterfactual Time & Counterfactual rank \\
\midrule
Hoey Indoor 800m WR  & 1:42.50 & Traditional one-turn indoor 800m & 0.08--0.13 & 1:42.58--1:42.63 & 1st (WR retained) \\
Kessler Indoor 2000m WR & 4:48.79 & \makecell[l]{Hocker '26 waterfall\\Hocker '24 back group} & 0.42--1.31 & 4:49.21--4:50.10 & 1st--2nd all-time \\
Attaoui Indoor 1000m ER  & 2:14.52 & \makecell[l]{Hocker '26 waterfall\\Hocker '24 back group} & 0.58--1.39 & 2:15.10--2:15.91 & 2nd European mark \\
Hocker Indoor Mile AR & 3:45.94 & \makecell[l]{Hocker '26 waterfall\\Hocker '24 back group} & 0.63--1.49 & 3:46.57--3:47.43  & 1st (AR retained) \\
Wanyonyi Outdoor 1000m WR & 2:11.83  & Ngeny '99 & 0.39--0.42 & 2:12.22--2:12.25 & 3rd all-time \\
Kerr Outdoor Mile WR
& 3:42.66 & \makecell[l]{Ingebrigtsen '23\\Myers '26\\El Guerrouj '99} & 0.65--0.88 & 3:43.31--3:43.54 & 2nd--3rd all-time \\

\bottomrule
\end{tabular}
\end{center}

Applying the combined geometric and path-length model to the six 2026
performances produced estimated advantages ranging from 0.08s (lower bound for
Hoey's indoor 800m) to 1.49s (upper bound for Hocker's indoor mile)
(Table~\ref{tab:combined_results}). The geometric component was generally
small, approximately 0.02--0.06s in the outdoor comparisons and generally
below 0.15s indoors. The largest estimated advantages instead occurred in
comparisons where athletes covered substantial additional distance, with the
path-length component exceeding 1s in comparisons with Hocker's 2024
back-group performance. The code used to implement these calculations 
also allows any combination of realized distance and effective
lane position to be evaluated directly (Appendix: ~\nameref{asec:code}).

For the two outdoor world records, correcting for the starting configuration
changes their position on the all-time list. Kerr's estimated advantage was
0.65--0.88s, giving a counterfactual mile of 3:43.31--3:43.54 rather than
3:42.66. This places the performance solidly behind the previous WR
(El Guerrouj's 3:43.13, 1999) and, depending on the comparison, just ahead of
or behind Noah Ngeny, who took second behind El Guerrouj in the same race.
Wanyonyi's estimated advantage was 0.39--0.42s, moving his 2:11.83 1000m to
2:12.22--2:12.25. The corrected performance falls behind both Noah Ngeny's
former record (2:11.96, 1999) and Sebastian Coe (2:12.18, 1981), making it
the third-fastest performance all-time.

The indoor results were more dependent on the comparison start because the
difference in distance traveled was larger. 
Kessler's 4:48.79 indoor 2000m
becomes 4:49.21--4:50.10, placing it between first and second all-time
depending on the comparison. Attaoui's 2:14.52 European record becomes
2:15.10--2:15.91, which would make it the second-fastest European mark.
Hocker's 3:45.94 American record becomes 3:46.57--3:47.43 under the
counterfactual starting conditions.
Although the upper end of this range is
slower than Nuguse's uncorrected 3:46.63, Nuguse's mark was itself set from
the front group of a one-turn split-staggered start and included approximately
1.0m of excess distance. Applying the same comparison conditions to Nuguse
gives counterfactual times of 3:47.08--3:47.12 under the waterfall comparison
and 3:47.92--3:47.94 under the back-group comparison. Hocker therefore retains
the American record under both matched counterfactuals.
Hoey's indoor 800m is different because the comparison involves the starting
geometry alone rather than additional distance traveled. His estimated
advantage was only 0.08--0.13s, giving a counterfactual time of
1:42.58--1:42.63. Even with this correction, his performance remains the
fastest indoor 800m performance.

Taken together, these counterfactuals do not suggest that the exceptional
2026 season was simply an artifact of the split-staggered start. Even after
removing the estimated starting advantage, several of these performances
would still have rewritten the record books. Hoey retains the world record,
Hocker retains the American record, and Kerr, although falling behind
El Guerrouj's world record, still produces the fastest European mile.
Kessler either retains the world record or falls only to second all-time,
depending on the comparison used, while Attaoui remains the second-fastest
European ever and Wanyonyi the third-fastest athlete ever over 1000m.
Thus, 2026 would still have been an extraordinary year for middle-distance
running without the split-staggered start. The start does not explain the
underlying concentration of historically fast performances. Rather, its
importance appears at the margins of the record book: advantages of a few
tenths to roughly a second can determine whether an already exceptional
performance becomes a world or area record. The recent assault on the
middle-distance record books therefore cannot be attributed solely to the
starting configuration, but our results suggest that it has played a
meaningful role in determining which of those performances became records.

%%%%%%%%%%%%%%%%%%%%%%%%%%%%%%%%%%%%%%%%%%%%%%%%%%%%%%%%%%%%%%%%%%%%%%%%%%%%%%%%%%%%%%%%%%%%%%%  
%%%%%%%%%%%%%%%%%%%%%%%%%%%   DISCUSSION %%%%%%%%%%%%%%%%%%%%%%%%%%%%%%%%%%%%%%%%%%%%%%%%%%%%%%  
%%%%%%%%%%%%%%%%%%%%%%%%%%%%%%%%%%%%%%%%%%%%%%%%%%%%%%%%%%%%%%%%%%%%%%%%%%%%%%%%%%%%%%%%%%%%%%%  

\section*{Discussion}
\label{sec:discussion}

The main results of this manuscript demonstrate that the split-staggered start is not time-neutral: 
placement in the front group provides a modest but meaningful advantage to athletes. 
It does not explain why these athletes were capable of historically
exceptional performances or suggest that 2026 would not have been an extraordinary year
for middle-distance running without it. But record margins are small, and an
advantage of a few tenths of a second can determine whether a record-caliber
performance actually becomes a record. There may, however, be a more
consequential effect of the split stagger that our analysis cannot fully
capture: athletes in the back group may be disadvantaged relative not only to
the front group, but also to athletes starting from a conventional waterfall.

The geometry of the merge provides a plausible reason. Until the groups
converge, back-group athletes run the tighter turn, while the front group
runs at a wider radius. At the merge, the front group is also moving inward
and, indoors, down the bank, while the back group must either yield position or
accelerate around athletes arriving from outside, generally while covering
additional distance. Our model captures the distance cost but not the cost of
accelerating, yielding position, or defending the inside as another group
merges into its path. Cole Hocker's race history provides some suggestive examples: 
he ran approximately 4m extra from a traditional waterfall start in his 
2026 indoor 2000m, compared with approximately 10m when starting in the back group at Millrose in 2024.
Nuguse similarly ran 3:43.97 in the mile from a waterfall start in 2023 compared to 
3:45.69 from the back group when he finished second to Kerr in 2026. 

These comparisons cannot isolate the effect of the start, but they raise the 
possibility that a traditional waterfall was relatively even: starting outside 
imposed a distance cost, while starting inside imposed a tactical cost. A split stagger may alter
that balance in both directions, giving the smaller front group an advantage
while imposing an additional disadvantage on the larger back group. If so,
the more important consequence is competitive rather than historical. The
starting procedure could make it harder for an unexpectedly strong athlete to
reach and defeat the favored athletes placed in front. 
That all 14 world records contested under the split-staggered start format were set from the
front group is consistent with this hypothesis. Whether major races and
records will continue to be won almost exclusively from the front group
may therefore be a more important test of
the format than the relatively small time advantages estimated here.

An interesting implication of these results is that they may help explain the
longstanding apparent weakness of the mile world record relative to the 1500m.
Simply extrapolating an additional 109.344m from the 1500m implicitly treats
the mile as a longer version of the same race. Under a traditional waterfall
start, however, the additional distance also changes where the race begins:
the 1500m begins near the start of a straight, whereas the mile begins before
a turn. Prior to his 2026 attempt, Kerr called the mile ``the hardest world
record right now'' and included starting on a bend among several reasons for
its difficulty \cite{kerr2026letsrun,letsrun2026kerr}. Yet the World Athletics
scoring tables rated El Guerrouj's 3:43.13 mile as a weaker performance than
his 3:26.00 1500m world record \cite{worldathletics2025scoring}. Kerr's record
provides an even more direct illustration: he passed 1500m in 3:27.62 during
his 3:42.66 mile, already 1.62s slower than the 1500m world record with another
109.344m still to run \cite{worldathletics2026kerrmile}. Our results suggest
that starting geometry may explain part of this discrepancy. Under a
traditional waterfall, the mile begins immediately before a turn, introducing
geometric and path-length costs that are largely absent from a 1500m record
attempt. The split-staggered start used in Kerr's successful attempt
substantially reduced precisely these costs.

The split stagger also illustrates a broader problem that arises when practical
changes to competition rules affect performance. World Athletics' recent
proposal replacing the long-jump take-off board with a camera-measured
take-off zone provides a useful comparison \citep{sportskeeda2024longjump}.
Both changes have straightforward practical justifications. A take-off zone
would reduce the number of fouls in the long jump, while a split-staggered
start can reduce crowding and collisions and allow larger fields, particularly
indoors. But just as the split stagger can reduce geometric and tactical costs 
and allow athletes to avoid running extra meters at the start of the race,
the take-off zone would eliminate centimeters lost due to the challenge 
athletes face staying behind the board. 

This tradeoff matters because better performances and records are themselves
valuable to many of the stakeholders in elite athletics. Sebastian Coe was
unusually explicit about this tension when discussing the proposed long-jump
change, arguing that athletics had to consider the interests of spectators and
sponsors as well as those of athletes \citep{sportskeeda2024longjump}.
Athletes including Olympic champion Miltiadis Tentoglou objected that accurately
attacking the board was part of the event, and World Athletics ultimately
withdrew the proposal in 2025
\citep{sportskeeda2024longjump,insidethegames2025longjump}.
The performance effect of the split stagger is considerably more modest and
less obvious than the proposed change to the long jump, which may help explain
why it has attracted much less scrutiny. But the underlying issue is similar:
a rule change can have a legitimate practical justification while also making
better marks easier to produce.

There is therefore little reason to expect athletes, coaches, agents, or meet
organizers to ignore a legal starting configuration that appears favorable for
a record attempt. If one configuration is faster, using it is a rational
response to the rules rather than evidence of wrongdoing. The same search for
marginal gains occurs through pacing, equipment, track design, and race
organization. The narrower conclusion of the present study is that starting
configuration belongs on this list. Split-staggered and waterfall starts
should not be assumed to be physically equivalent simply because both are
legal, and differences in starting configuration should be considered when
comparing performances across races and eras.

\renewcommand{\tablename}{\textbf{Appendix Table}}
\setcounter{table}{0}
\renewcommand{\theHtable}{appendix.\arabic{table}}
\clearpage \newpage 
\appendix
\part*{Appendix}
\vspace{1cm} 
\section*{Middle-Distance Records} 
\label{asec:records} 
\vspace{0.2cm} 
Appendix Table~\ref{atab:record_timeline} lists all men's middle-distance world-record
performances since 2000, together with the start configuration and date
of each. Fourteen of the twenty-four listed performances occurred between 2022 and 2026.
Under a constant-rate null over the full 2000--2026 window, the probability that 14 or
more of 24 performances would fall within the final 5 of 27 years is
$p \approx 1.7\times10^{-5}$, from the exact binomial tail probability
$P(X \geq 14) = 1.656\times10^{-5}$ for $X \sim \mathrm{Binomial}(24,\, 5/27)$.

Of the fourteen split-staggered-start records listed in Table~\ref{atab:record_timeline},
all fourteen were set from the forward group. Under the deliberately naive null that
record setters are drawn at random from the field, with the forward group comprising
approximately one-third of competitors, the probability of all 14 coming from the
forward group is $(1/3)^{14} \approx 2.1\times10^{-7}$, or roughly one in five million.
This calculation assumes a constant one-third forward-group share and independence
across performances; neither assumption is likely to hold exactly, and the true forward-
group fraction varies somewhat by race (Appendix Table~\ref{atab:path_inputs}). The
figure is intended as a naive baseline against which to judge the observed pattern, not
as a precise estimate of its probability under a realistic null.
\begin{center}
\small
\captionof{table}{Mid-Distance World Records/Bests Since 2000.} \label{atab:record_timeline}
\begin{tabular}{llllll}
\toprule
Event & Rank & Time & Athlete & Start Configuration & Date \\
\midrule
2\,mile Indoor  & 5 & 8:04.69 & Haile Gebrselassie & Traditional Waterfall & 21.02.2003 \\
2\,mile Indoor  & 4 & 8:04.35 & Kenenisa Bekele & Traditional Waterfall & 16.02.2008 \\
2000\,m Indoor  & 3 & 4:49.99 & Kenenisa Bekele & Traditional Waterfall & 17.02.2007 \\
800\,m Outdoor  & 3 & 1:41.09 & David Rudisha & 1-Turn Lane-Locked & 22.08.2010 \\
800\,m Outdoor  & 2 & 1:41.01 & David Rudisha & 1-Turn Lane-Locked & 29.08.2010 \\
800\,m Outdoor  & 1 & 1:40.91 & David Rudisha & 1-Turn Lane-Locked & 09.08.2012 \\
2\,mile Indoor  & 2 & 8:03.40 & Mohamed Farah & Traditional Waterfall & 21.02.2015 \\
1000\,m Indoor  & 1 & 2:14.20 & Ayanleh Souleiman & 1-Turn Split Stagger (F) & 17.02.2016 \\
1500\,m Indoor  & 4 & 3:31.04 & Samuel Tefera & 1-Turn Split Stagger (F) & 16.02.2019 \\
Mile Indoor     & 4 & 3:47.01 & Yomif Kejelcha & 1-Turn Split Stagger (F) & 03.03.2019 \\
1500\,m Indoor  & 2 & 3:30.60 & Jakob Ingebrigtsen & 1-Turn Split Stagger (F) & 17.02.2022 \\
3000\,m Indoor  & 3 & 7:23.81 & Lamecha Girma & 1-Turn Split Stagger (F) & 15.02.2023 \\
2\,mile Outdoor & 1 & 7:54.10 & Jakob Ingebrigtsen & Traditional Waterfall & 09.06.2023 \\
2000\,m Outdoor & 1 & 4:43.13 & Jakob Ingebrigtsen & Traditional Waterfall & 08.09.2023 \\
2\,mile Indoor  & 1 & 8:00.67 & Josh Kerr & 1-Turn Split Stagger (F) & 11.02.2024 \\
3000\,m Outdoor & 1 & 7:17.55 & Jakob Ingebrigtsen & 1-Turn Split Stagger (F) & 25.08.2024 \\
3000\,m Indoor  & 1 & 7:22.91 & Grant Fisher & 1-Turn Split Stagger (F) & 08.02.2025 \\
Mile Indoor     & 3 & 3:46.63 & Yared Nuguse & 1-Turn Split Stagger (F) & 08.02.2025 \\
1500\,m Indoor  & 1 & 3:29.63 & Jakob Ingebrigtsen & 1-Turn Split Stagger (F) & 13.02.2025 \\
Mile Indoor     & 1 & 3:45.14 & Jakob Ingebrigtsen & 1-Turn Split Stagger (F) & 13.02.2025 \\
800\,m Indoor   & 1 & 1:42.50 & Josh Hoey & 2-Turn Lane-Locked & 24.01.2026 \\
2000\,m Indoor  & 1 & 4:48.79 & Hobbs Kessler & 2-Turn Split Stagger (F) & 24.01.2026 \\
1000\,m Outdoor & 1 & 2:11.83 & Emmanuel Wanyonyi & 1-Turn Split Stagger (F) & 10.07.2026 \\
Mile Outdoor    & 1 & 3:42.66 & Josh Kerr & 1-Turn Split Stagger (F) & 18.07.2026 \\
\bottomrule
\end{tabular}
\end{center}

\clearpage \newpage 
\section*{Data and Code Availability}
\label{asec:code} 
The code used to perform the geometric, path-length, and counterfactual
calculations reported in this study is publicly available at
\url{https://github.com/tadesouaiaia/trackTools}.
The race-distance measurements and other inputs used in the analyses are
reported in the Appendix and their original sources are cited in the text.

\section*{Track Constants} 
\label{asec:constants} 
\vspace{0.2cm} 
The following table lists the track constants used in this manuscript, from the World Athletics Manual \citep{worldathletics2008}. 
Lane $n$ running-line radius calculated as $r_n=r_1+(n-1)w$.  Note that the World Athletics Manual lists the standard 
indoor lane-width at 1.0m, but the four records analyzed in 2026 were set at modern tracks \cite{new_balance_track, sound_running, gallur_madrid} 
which use the maximal allowable width of 1.07m per lane. 
\vspace{1cm} 
\begin{center}
\captionof{table}{Track Constants}
\label{atab:constants}
\begin{tabular}{lcc}
\toprule
& Outdoor 400 m & Indoor 200 m \\  \midrule
Radius, lane 1 (running line), $r_1$ & 36.80 m & 17.496 m \\
Straight length (each) & 84.39 m & 35.688 m \\
Curve arc length, one turn, lane 1 ($L_q$) & 115.61 m & 64.312 m \\
Lane width, $w$ & 1.22 m & 1.00 m (calibration); 1.07 m (2026 races) \\ \bottomrule
\end{tabular}
\end{center}

\clearpage  
\section*{Effective Path Length and Lane Position for Recent Records and Comparators}
\label{asec:paths} 
\vspace{0.2cm} 
Two quantities are reported in Table~\ref{atab:path_inputs} for each performance:
effective lane, used in the Factor~1 geometric correction, and extra distance, used in
the Factor~2 path-length correction. Comparator performances were selected using the
criterion stated in Methods, restricted to races run at genuine record pace, where the
tactical costs of inside and outside positioning described above are in force.
Five comparator performances are record attempts in the strict sense: El Guerrouj's 1999 mile
world record; Ingebrigtsen's 2023 Bowerman Mile, a failed world-record attempt that set a
European record; Myers's 2026 Bowerman Mile, a continental (Oceanian) record; Ngeny's 1999
world record 1000m; and Hocker's 2026 American record 2000m, set from an indoor
waterfall start.

We were unable to identify any examples of winning athletes setting area or world records from 
the back group in a split-stagger start. Whether this is because real challenges reduce 
the likelihood of winning from the back or because perceived benefits lead meet organizers 
to place pre-race favorites and their pace-setters in the front group, this makes the 
counterfactual of how fast these performances would have been from a back-group starting position hard to answer directly.
The closest available case is Hocker's 2024 Millrose 2-mile, in which Hocker (an Olympic and world champion, featured in this 
analysis) started in the back group in a race where his front group rival Josh Kerr broke the indoor 2-mile record. 
Hocker ran an estimated 10m extra distance across two separate detours - an initial 
attempt to work around the back group and again when forced wide by the front group 
merging down the bank.  While an imperfect single data-point, this race can be read 
alongside Hocker's record setting indoor waterfall performance - the true back-group 
cost for a genuine favorite plausibly lies between the two. Hocker is, to our knowledge, the only
contemporary elite athlete (World/Olympic Champion in 2024, 2025, 2026) to have competed 
in record setting indoor races in all three relevant positions (Waterfall, Front and Back group in split-stagger starts), 
thus, his results suitably represent the counterfactual for indoor racing. 

Hoey's 800\,m has no comparator performance in Table~\ref{atab:path_inputs}, his novel 
race configuration is a two-turn vs a traditional one-turn multi-lane stagger - neither 
is associated with an extra distance burden at the start of the race.  Thus, his advantage 
is purely geometric - the benefit of an extra one hundred meters along the outside before 
descending the bank and running along the rail, there is no path length component included. 

\begin{center}
\captionof{table}{Lane-position and path-length measurements for the 2026 record
performances and comparison races. Effective lane is the mean lane position
over the opening turn for one-turn configurations and the opening two turns
for two-turn configurations. Extra distance is measured over the full race.
SS = split-stagger, WF = waterfall; F/B indicate forward/back group.
Video and OMEGA extra-distance estimates are reported separately where available.
Basis establishes the record-setting nature of the race: WR=World Record, AR=Area Record (Continental).} 
\label{atab:path_inputs}
\small
\begin{tabular}{llcccll}
\toprule
Athlete & Race & Config. & \shortstack{Eff.\\Lane} & \shortstack{Extra\\(Video)} & \shortstack{Extra\\(OMEGA)} & Basis \\
\midrule
Hoey & Indoor 800m, 2026 & 2-turn lane-stagger & 5.0 & 0m & -- & WR \\
Kessler & Indoor 2000m, 2026 & 2-turn SS (F) & 4.0 & 1.5m & -- & WR \\
Attaoui & Indoor 1000m, 2026 & 2-turn SS (F) & 4.0 & 0.25m & -- & AR \\
Hocker & Indoor mile, 2026 & 2-turn SS (F) & 4.0 & 0m & -- & AR \\
Wanyonyi & Outdoor 1000m, 2026 & 1-turn SS (F) & 5.0 & 0.5m & 1.0m \citep{OmegaMonaco1000m2026} & WR \\
Kerr & Outdoor mile, 2026 & 1-turn SS (F) & 5.0 & 0m & 0.0m \citep{omega_london2026} & WR \\
\midrule \multicolumn{7}{c}{\textit{Comparators}} \\ \midrule
El Guerrouj & Rome Mile, 1999 & WF & 2.3 & 6.0m & -- & WR \\
Ingebrigtsen & Bowerman Mile, 2023 & WF & 2.1 & 4.5m & 4.56m \citep{OmegaBowerman2023} & WR Attempt; AR \\
Myers & Bowerman Mile, 2026 & WF & 2.4 & 5.5m & 5.46m \citep{OmegaBowerman2026} & AR \\
Ngeny  & Rome 1000m, 1999 & WF & 2.0 & 3.75m & -- & WR \\
Hocker & Hokie 2000m, 2026 & WF & 2.0 & 4m & -- & AR \\
Hocker & Millrose 2 Mile, 2024 & 1-turn SS (B) & 2.7 & 10m & -- & Kerr WR (Hocker 3rd)   \\
\bottomrule
\end{tabular}
\end{center}
\clearpage
\bibliographystyle{unsrtnat} 
\bibliography{refs}

@misc{new_balance_track,
      author       = {{The TRACK at New Balance}},
        title        = {The TRACK at New Balance: Facility Specifications and Track Dimensions},
          howpublished = {\url{https://thetrackatnewbalance.com}},
            note         = {Accessed: 2026-09-09},
              year         = {2026}
}

@misc{sound_running,
      author       = {{Sound Running}},
        title        = {Sound Running Event Specifications and Indoor Invite Formats},
          howpublished = {\url{https://soundrunning.run}},
            note         = {Accessed: 2026-09-09},
              year         = {2026}
}

@misc{gallur_madrid,
      author       = {{Real Federaci\'{o}n Espa\~{n}ola de Atletismo}},
        title        = {Centro Deportivo Municipal Gallur: Madrid Indoor Meeting and Venue Specifications},
          howpublished = {\url{https://rfea.es}},
            note         = {Accessed: 2026-09-09},
              year         = {2026}
}

@article{crystalpalace,
      author  = {{World Athletics}},
        title   = {Past Meets Present as Crystal Palace Prepares for World Mile Record Attempt},
          year    = {2000},
            month   = aug,
              day     = {4},
                journal = {World Athletics},
                  url     = {https://worldathletics.org/news/news/past-meets-present-as-crystal-palace-prepares},
                    note    = {Accessed 8 September 2026}
}

@article{rome,
      author  = {Landells, Steve},
        title   = {Fab Five: Mile Races},
          journal = {World Athletics},
            year    = {2019},
              month   = jul,
                url     = {https://worldathletics.org/news/series/fab-five-mile-races},
                  note    = {Accessed 8 September 2026}
}

@article{ingebrigtsen2023mileattempt,
      author  = {{World Athletics}},
        title   = {2023 Review: Middle and Long Distance},
          journal = {World Athletics},
            year    = {2023},
              url     = {https://worldathletics.org/news/series/2023-review-middle-long-distance},
                note    = {Accessed 8 September 2026}
}

@article{morceli1993mile,
      author  = {{The Washington Post}},
        title   = {Algeria's Morceli Sets World Record in Mile},
          journal = {The Washington Post},
            year    = {1993},
              month   = sep,
                day     = {5},
                  url     = {https://www.washingtonpost.com/archive/sports/1993/09/06/algerias-morceli-sets-world-record-in-mile/1cde3fa1-ba8f-4bed-ae28-dfbd3581ee55/},
                    note    = {Accessed 8 September 2026}
}

@article{cram1985mile,
      author  = {{World Athletics}},
        title   = {40 Years On -- Revisiting the World Records of Kristiansen, Aouita and Cram},
          journal = {World Athletics},
            year    = {2025},
              month   = jul,
                day     = {27},
                  url     = {https://worldathletics.org/heritage/news/40-years-on-world-records-kristiansen-aouita-cram-oslo},
                    note    = {Accessed 8 September 2026}
}

@book{worldathletics2025scoring,
      author    = {{World Athletics}},
        title     = {World Athletics Scoring Tables of Athletics},
          year      = {2025},
            edition   = {Revised Edition},
              note      = {By Bojidar Spiriev; updated by Attila Spiriev},
                publisher = {World Athletics},
                  url       = {https://worldathletics.org/about-iaaf/documents/technical-information}
}

@article{worldathletics2026kerrmile,
      author  = {{World Athletics}},
        title   = {Mile World Record: How Kerr Added His Name to the History Book},
          journal = {World Athletics},
            year    = {2026},
              month   = jul,
                day     = {27},
                  url     = {https://worldathletics.org/news/feature/mile-world-record-josh-kerr-history-book}
}

@article{letsrun2026kerr,
      author  = {{LetsRun.com}},
        title   = {``My Race, My Rules'': Why Josh Kerr Is Chasing El Guerrouj's Mile World Record in London},
          journal = {LetsRun.com},
            year    = {2026},
              month   = apr,
                day     = {7},
                  url     = {https://www.letsrun.com/news/2026/04/my-race-my-rules-why-josh-kerr-is-chasing-el-guerroujs-mile-world-record-in-london/}
}

@misc{kerr2026letsrun,
      author       = {Kerr, Josh},
        title        = {Josh Kerr: Chasing Mile World Record + Beating Cole Hocker},
          howpublished = {LetsRun.com's Track Talk: The Home of Running and Track and Field},
            year         = {2026},
              month        = apr,
                day          = {5},
                  note         = {Interview with Jonathan Gault and Robert Johnson; discussion of mile versus 1500 m world-record difficulty begins at 28:50},
                    url          = {https://podcasts.apple.com/nz/podcast/josh-kerr-guest-chasing-mile-world-record-beating-cole/id383631335?i=1000759438704}
}

@article{kerr_shoes,
      author = {Johanna Gretschel},title = {Josh Kerr Broke the World Record in the Mile. His Custom Shoes Were Designed to Be a ``Weapon''},
          journal = {Runner's World},year = {2026},month = jul,day = {18},url = {https://www.runnersworld.com/news/a73176078/kerr-gear/}
}

@article{kerr_suit, 
author = {}, title = {Josh Kerr's Custom Brooks Suit and Spikes Are Tuned for Exactly 222 Seconds of All-Out Effort},
journal = {Believe in the Run},year = {2026},month = jul,day = {8},
  url = {https://believeintherun.com/josh-kerr-brooks-222-attempt/}
}

@article{kerr_chamber,
author = {Sean Abrams},
title = {The World's Best Miler Spends 12 Hours a Day Inside His Bedroom. Is Josh Kerr Crazy or Brilliant?},
journal = {Runner's World},year = {2026},month = jul,day = {18},
  url = {https://www.runnersworld.com/news/a71858125/josh-kerr-altitude-chamber-bedroom/}
}

@article{mar_shoes,
author = {Cameron Ormond},
title = {Adidas' Marathon Record Shoe Hits \$4,600 on Resale},
journal = {Canadian Running Magazine},
year = {2026},month = apr,day = {28},
  url = {https://runningmagazine.ca/the-scene/adidas-marathon-record-shoe-hits-4600-resale-prices/}
}

@article{bicarb,
author = {Jim Cotton},
title = {How Bicarb Became the Latest Must-Have of Endurance},journal = {Velo},
year = {2024},month = oct,day = {11},
  url = {https://velo.outsideonline.com/road/road-training/how-bicarb-became-the-latest-must-have-of-endurance/}
}

@article{michael_johnson,
author = {Jaime Davilla},
title = {Michael Johnson: "A los fans no les importa nada el debate de las zapatillas},
journal = {AS},
year = {},
month = {},
day = {},
  url = {https://as.com/masdeporte/atletismo/michael-johnson-a-los-fans-no-les-importa-nada-el-debate-de-las-zapatillas-n}
}

@article{no_shoes,
      author = {Wu, Y. and Zhang, H. and Wang, S. and Lu, C. and Xing, Q. and Tian, Y. and He, D. and Sun, L. and Shen, Y.},
        title = {Comparative analysis of foam-only versus carbon-plated advanced footwear technology spikes in distance runners},
          journal = {Frontiers in Physiology},
            year = {2025},
              volume = {16},
                pages = {1703854},
                  doi = {10.3389/fphys.2025.1703854},
                    pmid = {41347023},
                      pmcid = {PMC12672227},
                        url = {https://pmc.ncbi.nlm.nih.gov/articles/PMC12672227/}
}

@misc{FlashResults2026Hokie2000m,
  author = {{Flash Results}},
  title = {Hokie Invitational 2026: Men's 2000 m Final Results},
  year = {2026},
  url = {https://www.flashresults.com/2026_Meets/Indoor/01-23_HokieInvite/044-1_compiled.htm},
  note = {Accessed: 2026-09-07}
}

@misc{Monti2026HoeyKessler,
  author = {Monti, David},
  title = {Hoey, Kessler Run World Records at New Balance Indoor Grand Prix 2026},
  year = {2026},
  url = {https://www.flotrack.org/articles/15293955-hoey-kessler-run-world-records-at-new-balance-indoor-grand-prix-2026},
  note = {Accessed: 2026-09-07}
}

@techreport{worldathletics2008,
  author      = {{World Athletics (formerly IAAF)}},
  title       = {Track and Field Facilities Manual, 2008 Edition -- Marking Plans (200m Indoor Track and 400m Outdoor Track)},
  institution = {World Athletics},
  year        = {2008}
}

@article{munro2022,
  author  = {Munro, David},
  title   = {Are There Lane Advantages in Track and Field?},
  journal = {PLOS ONE},
  year    = {2022},
  volume  = {17},
  number  = {8},
  pages   = {e0271670},
  doi     = {10.1371/journal.pone.0271670}
}

@misc{chen2022blog,
  author = {Chen, Jeff},
  title  = {Effect of Lane Draw in 200m Sprinters},
  year   = {2022},
  note   = {Blog post, wind- and season-best-corrected regression on Diamond League data},
  url    = {https://jeffchen.dev/posts/Effect-of-Lane-Draw-In-200m-Sprinters/}
}

@inproceedings{flanders_isbs,
      author    = {Delecluse, Christophe and Suy, E. and Diels, R.},
        title     = {Effect of Curve and Slope on Indoor Track Sprinting},
          booktitle = {Proceedings of the 16th International Symposium on Biomechanics in Sports (ISBS 1998)},
            year      = {1998},
              pages     = {149--152},
                address   = {Konstanz, Germany},
                  url       = {https://ojs.ub.uni-konstanz.de/cpa/article/view/959},
                    note      = {Flanders Expo indoor track, Ghent, Belgium; host venue of the 2000 European Indoor Championships}
}

@article{chang1999,
  author  = {Chang, Young-Hui and Kram, Rodger},
  title   = {Metabolic Cost of Generating Horizontal Forces During Human Running},
  journal = {Journal of Applied Physiology},
  year    = {1999},
  volume  = {86},
  number  = {5},
  pages   = {1657--1662}
}

@article{chang2007,
  author  = {Chang, Young-Hui and Kram, Rodger},
  title   = {Limitations to Maximum Running Speed on Flat Curves},
  journal = {Journal of Experimental Biology},
  year    = {2007},
  volume  = {210},
  number  = {6},
  pages   = {971--982},
  doi     = {10.1242/jeb.02728}
}

@article{churchill2016,
  author  = {Churchill, Sarah M. and Trewartha, Grant and Bezodis, Ian N. and Salo, Aki I. T.},
  title   = {Force Production During Maximal Effort Bend Sprinting: Theory vs Reality},
  journal = {Scandinavian Journal of Medicine \& Science in Sports},
  year    = {2016},
  volume  = {26},
  pages   = {1171--1179},
  doi     = {10.1111/sms.12559}
}

@misc{OmegaBowerman2023,
  author       = {{OMEGA Timing}},
  title        = {The Bowerman Mile Men: Race Analysis -- Graphical},
  year         = {2023},
  howpublished = {Wanda Diamond League, Eugene},
  url          = {https://www.omegatiming.com/File/000203100102090101FFFFFFFFFFFF4E.pdf},
  note         = {Official race analysis; includes Starting Order and Real Distance. Reports Ingebrigtsen's Real Distance as 1613.9\,m}
}

@misc{OmegaBowerman2026,
  author       = {{OMEGA Timing}},
  title        = {The Bowerman Mile Men: Race Analysis -- Graphical},
  year         = {2026},
  howpublished = {Wanda Diamond League, Eugene},
  url          = {https://www.omegatiming.com/File/0002060C01010C0101FFFFFFFFFFFF4E.pdf},
  note         = {Official race analysis; includes Starting Order and Real Distance. Reports Myers' Real Distance as 1614.8\,m}
}

@misc{omega_london2026,
  author = {{OMEGA Timing}},
  title  = {London 2026 -- Race Analysis: 1 Mile Men},
  year   = {2026},
  note   = {Official Wanda Diamond League race-analysis report, London Diamond League, 18 Jul 2026. Reports Kerr's Real Distance as 1609.0\,m},
  url    = {https://www.omegatiming.com/File/0002060E01020C0101FFFFFFFFFFFF4E.pdf}
}

@misc{OmegaMonaco1000m2026,
  author = {{OMEGA}},
  title = {Herculis Monaco 2026: Men's 1000 m Race Analysis},
  year = {2026},
  url = {https://www.omegatiming.com/File/0002060D0102040101FFFFFFFFFFFF4E.pdf},
  note = {Graphical race analysis; accessed 2026-09-07}
}

@misc{sportskeeda2024longjump,
      author       = {{Sportskeeda}},
        title        = {``It's So Bad It's Funny'' -- {Olympic} Long Jump Champion {Miltiadis Tentoglou} Denounces {World Athletics} Proposal to Change the Long Jump to a Takeoff Zone},
          year         = {2024},
            url          = {https://sportskeeda.com/us/olympics/news-it-s-bad-funny-olympic-long-jump-champion-miltiadis-tentoglou-denounces-world-athletics-proposal-change-long-jump-takeoff-zone},
              note         = {Includes Sebastian Coe's statement that rule changes are necessary to ``keep up with the evolving interests of sponsors and viewers''}
}

@misc{insidethegames2025longjump,
      author       = {{Inside the Games}},
        title        = {{World Athletics} Rules Out Takeoff Zone},
          year         = {2025},
            month        = dec,
              url          = {https://www.insidethegames.biz/articles/1156056/world-athletics-rules-out-takeoff-zone},
                note         = {CEO Jon Ridgeon confirms withdrawal of the long jump takeoff-zone proposal following athlete opposition}
}
\end{document}